# A METHOD FOR COMPARING CHESS OPENINGS[1]

JAMAL MUNSHI

ABSTRACT. A quantitative method is described for comparing chess openings. Test openings and baseline openings are run through chess engines under controlled conditions and compared to evaluate the effectiveness of the test openings. The results are intuitively appealing and in some cases they agree with expert opinion. The specific contribution of this work is the development of an objective measure that may be used for the evaluation and refutation of chess openings, a process that had been left to thought experiments and subjective conjectures and thereby to a large variety of opinion and a great deal of debate.[2]

## 1. INTRODUCTION

The first few moves in chess are referred to collectively as the opening. The importance of the opening is widely recognized. Its importance rests on the assumption that it affects the outcome of the game. The rationale for this assumption is that in the opening both white and black attempt to deploy their pieces and simultaneously to occupy and control the four empty squares in the middle of the board and the party that gets ahead in achieving these goals enjoys a higher probability of winning the game ceteris paribus (Horowitz, 1964)

Chess openings have been an intense area of research for more than four hundred years in the European era of the game and many hundreds of standardized openings are now recognized, organized, and listed according to a system known as the Encyclopedia of Chess Openings or ECO (Wikibooks, 2013). These opening moves evolved over the centuries from the analyses and games of the grandmasters of chess. All serious chess players today begin their games with one of these proven opening sequences.

These openings differ from one another with respect to how well they facilitate their objective of center control and piece deployment. A direct comparison of openings in this regard is made difficult by the complexity of their intended function. Openings with a weakness in one aspect of the game may have a compensating strength in another or those that are strong in the beginning may offer opportunities for a counter attack later in the game. Also, openings that don't control the center with pawns and knights may do so remotely with bishops; or an apparent opportunity for the opposition may serve as a trap (Horowitz, 1964). These complexities have made it difficult to compare openings directly or to refute weak openings because no universal objective measure exists to make the necessary comparisons (Chessvibes, 2009). To judge the relative merit of an opening, chess players depend on two sources - grandmaster analyses and game statistics. Both of these resources are unreliable.

The grandmasters of the game are able to look ahead ten moves or more to assess all the possible ways the game can evolve from a given position. They use these faculties to make their picks for strong openings and to declare their refutations for weak ones. In fact, we owe the existence of rote "book" openings to these kinds of analyses that have been published since the 17th century (Horowitz, 1964). Yet there is no general agreement among the grandmasters on this issue even to the point that we may find that an opening that has been refuted by one grandmaster is actually played by another. For example, the King's Gambit, famously refuted by former world champion Bobby Fischer (Fischer, 1961), continues to be played even by grandmasters (chessgames.com, 2014) Thus, grandmaster analyses by themselves, though necessary, are not sufficient to assess the relative merits of chess openings.

The second resource for comparing openings are large databases containing the first ten or so moves from each of millions of games organized into an "Opening Book" format (Meyer-Kahlen, 2007). These databases allow the user to view the popularity as well as win, loss, and draw statistics for each opening line. Although universally used, these data are not as useful as they might appear because they are field data that have not been taken under controlled conditions. They are confounded by larger effects than the effect of the opening that they profess to measure.

Two important uncontrolled variables in these data are (1) the level of play and (2) the difference in playing strength between the two players. These variables are likely to have a greater effect on game outcome than the choice of opening. The lower the level of play, the higher the move imperfection rate, and the greater the effect of imperfect moves on game outcome. The effect of the opening on game outcome may not be detected under these conditions. Likewise, the greater the difference in playing strength between the two opponents the greater will be its effect on game outcome with the relatively weaker effect of the opening not observable under these conditions. The variance in game outcomes created by these stronger explanatory variables makes it difficult to measure the effect of the opening. Opening effectiveness data taken under these conditions may be unable to discriminate between good and bad openings. This work proposes a method for generating chess game outcomes in controlled experiments designed to detect the effect of the opening on game outcome.

## 2. THEORY

A chess game is a trinomial stochastic process. The three possible outcomes of a game may be enumerated as white wins, black wins, and draw and so we may model the chess game as a discrete stochastic process driven by an unknown and unobservable probability vector (Wikipedia, 2013) with two degrees of freedom expressed as

$$\pi[p_w, p_b, p_d]$$

where $\pi$ is the probability vector that generates game outcomes, $p_w =$ the probability that white will win, $p_b =$ the probability that black will win, and $p_d = 1 - p_w - p_b =$ the probability that the game will end in draw. As $p_w$ and $p_b$ cannot be assumed to be independent, it is necessary treat them as a vector and to analyze them in two dimensional space. The Cartesian space xy defined as x= $p_w$ and y= $p_b$ contains all possible chess games. Each (x,y) point in this space represents a unique chess game.

For ease of evaluating any given opening, we assume that there exist neutral and perfect chess openings in which each side plays a move from a set of best possible moves so that neither side gains or cedes an advantage due to the opening. Before the game starts, white enjoys a first move advantage that can be translated into a certain probability $p_w$ =FMA that white will win (Wikipedia, 2013). Thus in the simplest case where the two players are both equal and perfect chess players who can correctly evaluate all possible future states of the board for every possible move, and who have played a perfect and neutral opening, the probability that the game will be decisive is FMA and the probability vector may be described with:

$$p_w = \text{FMA}, \; p_b = 0, \; p_d = 1\text{-FMA}$$
$$\pi[p_w, p_b, p_d]=[\text{FMA},0,1\text{-FMA}]$$

That is to say, perfect chess games end in draw except for a certain probability, likely to be small, that white will win because it makes the first move. Black has no chance of winning a perfect game. Now, suppose that the two players are equal but not perfect. Random imperfect moves in the games will create a certain probability of decisive games that applies equally to both sides. The probability of a decisive game is the sum of the two independent probabilities IMP=probability of imperfection and FMA=the first move advantage. Under these conditions, the win probabilities are as follows:

$$p_w = \text{FMA} + \text{IMP}/2, \; p_b = \text{IMP}/2, \; p_d = 1 - \text{FMA} - \text{IMP}$$
$$\pi[p_w, p_b, p_d]=[\text{FMA+IMP}/2,\text{IMP}/2,1\text{-FMA-IMP}]$$

Thus black wins only by virtue of imperfections and white wins not only by imperfection but also because it holds the first move advantage. For example, if the first move advantage is given by FMA = 4% and the imperfection rate is given by IMP = 10% then under the neutral opening condition for two players of equal playing strength, $p_w = 0.04 + 0.10/2 = 9\%$, and $p_b = 0.10/2 = 5\%$. In this case, game outcomes are generated in a stochastic process driven by the probability vector $\pi[p_w, p_b, p_d]=[0.09,0.05,0.86]$. In any given game the probability that white will win is 9%, the probability that black will win is 5%, the probability that the game will be decisive is 14%, and the probability that the game will end in draw is 86%. Now suppose that at a higher level of play, the imperfection is reduced to 2%. In that case $p_w = 5\%$, $p_b = 1\%$, and $p_d = 94\%$ and therefore $\pi[p_w, p_b, p_d]=[0.05,0.01,0.94]$. These relationships are consistent with the observation that at higher levels of play more games end in draw and of the decisive games, white wins more than black. The FMA is undetectable at high levels of IMP.

If the opening is not neutral and not perfect then the probability vector $\pi[p_w, p_b, p_d]$ will be changed by the opening. This idea forms the basic principle of this work. If the imperfect opening is an innovation by black its imperfection may increase $p_w$ or decrease $p_b$, or it may do both. Likewise, an imperfect innovation by white may decrease $p_w$ or increase $p_b$ or it may do both. The net result is that the opening imperfection will change the probability vector. To measure this effect we use controlled experiments and collect game outcome data under conditions in which other variables are fixed at levels at which that the relatively weaker effect of the opening may be detected.

## 3. THE PROPOSED METHODOLOGY

The proposed method of comparing openings makes use of a class of software referred to as "chess engines" to set up controlled experiments in which the effect of the opening may be detected. A chess engine is a computer program that selects a chess move to make in any given position of the pieces on the chessboard (Wikipedia, 2014). It constructs a tree of all possible move sequences[3] to a given search depth and calculates the positional value for each side at the end of each node of the tree and then selects the move that yields the greatest positional advantage. Typically millions of positions may be evaluated to make a single move. Although the engine's method of making moves is dramatically different from the way humans play chess it generates the same kind of game and, at sufficient depth, the engines can outplay humans even at the highest levels (Wikipedia, 2014).

To be able to detect the effect of the opening, it is necessary that the two players be of the same playing strength so that the effect of strength differential on game outcome is removed. In our engine experiment this condition is achieved by setting up an engine match in which both sides of the board are played by copies of the same engine utilizing identical parameters and set to evaluate board positions at the same search depth. The additional requirement is that the move imperfection rate should be low enough to render the effect of the opening detectable and measurable. This condition is realized by running the engines at a sufficient search depth of their look ahead move tree. The move imperfection rate IMP cannot be eliminated because at any finite search depth the positional values used to select a move may be inaccurate; but it is possible to hold IMP to a rate that is low enough for the effect of the opening to be be measured.

For our experiments we selected the Houdini3Prox64 chess engine, generally considered to be the industry leader (Wikipedia, 2013). The numeral "3" designates the version of the software, "64" designates that the software uses a 64-bit architecture, the word "Pro" indicates that more than six processors may be used. Version 4 of this software was released in November of 2013. It was not used in these experiments because it is relatively new and has not been tested in the field for a sufficient period of time. A preliminary exploratory study was used to select a search depth of 22 plies[4] and a sample size of 300 games per experiment for this study. A search depth of 22 plies, equivalent to looking ahead eleven moves, meets the criterion that the engine should look ahead at least ten moves. It is generally recognized that grandmasters use board pattern recognition in a way that is equivalent to a look-ahead of up to ten moves (Simon, 1996). A further condition for a relatively high level of play is that when using a standardized set of neutral openings, the imperfection rate, measured as black's win rate, should be low - and that the win rate of white should exceed that of black. Houdini3Prox64 playing at a search depth of 22 plies meets these conditions.

**3.1     Baseline and test openings.**  It is necessary to establish a baseline to serve as a control for chess game outcome statistics against which the performance of a test opening may be evaluated. The ideal baseline opening to use is one which is similar to the opening to be tested with a common stem in the move sequence so that the specific innovation of the test opening can be identified and evaluated. Ideally, the baseline opening should be neutral and "perfect" in the sense that each side responds with a move from the set of best possible moves and in course of the opening neither side cedes an advantage to the other due to the opening moves. To ensure that the outcome statistics are comparable, the number of half moves specified in the baseline is set equal to the half moves specified in the openings to be tested so that the point in the game where the engine begins to evaluate positional value is the same for the two openings being compared. A length of six half moves was selected and fixed for all openings used in the study.

---

[3] The tree is "pruned" to speed up the evaluation process
[4] One complete "move" in chess requires two "half moves" one from each side of the board. Chess engines measure their search depth in half moves usually referred to as "plies".

Baseline and test openings for this study are selected from a popular online opening database (Jones & Powell, 2014). This database allows the users to set the minimum Elo rating (Wikipedia, 2014) of the players when searching the database of chess games. For the purpose of this study we have restricted our search to a very high level of play by setting the minimum Elo rating to 2600. For the first three moves of chess games at this level we find that the most used opening is C68[5] Ruy Lopez[6] and the second most used opening line is B53 Sicilian Defense. We therefore select these lines as our baseline openings against which openings to be tested will be compared. Ten test openings are randomly selected for this study from a large set of openings from the same database that (1) share a common stem of initial moves with the baseline to which it will be compared and (2) are rarely played at the high level of play chosen. The selected baseline and test openings are shown in Table 1. The test openings are separated into two groups according to the baseline with which the test opening will be compared. Each test opening is identified by an "innovation" which marks the point of departure of the test opening from the stem it shares with the baseline. The relative *rarity* of the innovation is computed as a ratio of the frequency of the corresponding baseline move divided by the frequency of the innovation at the selected level of play. For example, the rarity of C50 Giuoco Piano[7] shown to be 9.0 means that at a level of play above 2600 Elo, the baseline move 3. Bb5 is played nine times more frequently than the innovation 3. Bc4. The word *innovation* carries no meaning other than to convey that it is a departure from the baseline.

| Expt# | Group | Opening Sequence | Description | Innovation | Ratio | Rarity |
|---|---|---|---|---|---|---|
| **1** | **1** | **e4e5 Nf3Nc6 Bb5a6** | **C68 Ruy Lopez** | **Baseline** | **3…a6/3…a6** | **1.0** |
| 2 | 1 | e4e5 Nf3Nc6 Bb5Nd4 | C61 Bird Defense | 3…Nd4 | 3…a6/3…Nd4 | 254.5 |
| 3 | 1 | e4e5 Nf3Nc6 Bc4Bc5 | C50 Giuoco Piano | 3. Bc4 | 3. Bb5/3. Bc4 | 9.0 |
| 4 | 1 | e4e5 Nf3Nc6 d4exd4 | C44 Scotch Game | 3. d4 | 3. Bb5/3. d4 | 10.9 |
| 5 | 1 | e4e5 Nf3d6 d4exd4 | C41 Philidor | 2…d6 | 2…Nc6/2…d6 | 100.2 |
| 6 | 1 | e4e5 f4exf4 Nf3g5 | C37 Kings Gambit | 2. f4 | 2. Nf3/2. f4 | 43.7 |
| **7** | **2** | **e4c5 Nf3d6 d4cxd4** | **B53Sicilian Defense** | **Baseline** | **2. Nf3/2. Nf3** | **1.0** |
| 8 | 2 | e4c5 d4cxd4 c3dxc3 | B21 Smith Morra | 2. d4 | 2. Nf3/2. d4 | 1459.2 |
| 9 | 2 | e4c5 c3d5 exd5Qxd5 | B22 Sicilian Alapin | 2. c3 | 2. Nf3/2. c3 | 35.8 |
| 10 | 2 | e4c6 d4d5 e5Bf5 | B12 Caro Kann | 1…c6 | 1…c5/1…c6 | 4.0 |
| 11 | 2 | e4d5 exd5Qxd5 Nc3Qd6 | B01 Scandinavian | 1…d5 | 1…c5/1…d5 | 39.6 |
| 12 | 2 | e4d6 d4Nf6 Nc3g6 | B07 Pirc | 1…d6 | 1…c5/1…d6 | 18.5 |

Table 1 Baseline and test openings selected for this study: UPDATED IN MARCH 2014

---

[5] The ECO designations apply to the first three specified moves. Transpositions are noted in the text.
[6] Also known as the Spanish Game
[7] Also known as the Italian Game

It is assumed that the innovator seeks a greater advantage than that offered by the neutral baseline opening sequence from which it has parted. There are three possible outcomes of an innovation. If the innovation is a success, we would expect it to change $\pi[p_w, p_b, p_d]$ in favor of the innovator. If the innovation is a failure, it backfires. In that case, we would expect it to change $\pi[p_w, p_b, p_d]$ to the detriment of the innovator. The third possibility is that the innovation fails to create the advantage sought by the innovator but it does no harm, in which case we would not expect to see any change in $\pi[p_w, p_b, p_d]$. We set up controlled experiments and hypothesis tests to select one of these effects as the likely outcome of the innovation.

Houdini 3x64 and Houdini3Prox64, products of cruxis.com (Houdanrt, 2012) are used for all engine experiments. The two engines are considered equivalent and differ only with respect to speed. The non-Pro version is limited to 6 processors (or cores) while the Pro version does not have such a limit. The engine parameters are set to the default values as shipped except for the number of threads which is set to the number of cores available for the Pro version and to 6 for the non-Pro version. All engine experiments in this study are carried out with contempt = 1. The Deep Shredder graphical user interface (GUI) from Shredder Chess (Meyer-Kahlen, 2007) was chosen for this study. Once the Houdini3 engine is installed in the GUI, we use the "Level of Play", "Opening Book", and "Engine Match" facilities of the GUI to set up each experiment. The Level of Play is set to a fixed search depth of 22 plies for each move the engines make. The Opening Book function is used to specify the opening to be used in the games by both sides. Finally, the Engine Match function is used to set up a match with either Houdini3Prox64 or Houdini3x64 playing both white and black moves. The number of games to be played is set to the selected sample size of 300 games. At the end of each match, the number of wins by white and the number of wins by black are recorded.

**3.2      Method for comparing match results.**  We assume that each game of an experiment is akin to tossing an unfair loaded three-sided "coin" in which the loadings are unknown. The loaded "coin" can come up as "white wins", "black wins", or "draw". The process is driven by an unknown probability vector $\pi[p_w, p_b, p_d]$.  In the controlled experiments the probability vector is assumed to be completely specified by three parameters, namely, white's first move advantage, the general rate of imperfection in the engine's position evaluation function, and the effect of the opening. The vector $\pi[p_w, p_b, p_d]$ is not known to us and it is not directly observable. It is necessary to infer its value from the data. The effectiveness of openings are then evaluated by comparing the inferred values of their respective probability vectors.

Each experiment of n=300 games is likened to tossing 300 of these three-sided unfair loaded coins. We then count w = the number of white wins and b = the number of black wins and compute

$$P[p_w, p_b, p_d] = [w/300, b/300, 1-w/300-b/300]$$

where $P[p_w, p_b, p_d]$ is an unbiased estimate of the unobservable probability vector $\pi[p_w, p_b, p_d]$. If we toss the coins again we are likely to get different values of w and b and therefore it is necessary to know how different these values could be. The uncertainty of the estimate is assessed using a simulation procedure to generate one thousand simulated repetitions of the 300-game experiment. The use of simulations to estimate variance is described by Grinstead and Snell (Snell, 1997). Microsoft Excel is used as the simulation tool. The procedure for generating simulated experiments is as follows. First we select a column of 300 random numbers from a uniform distribution with values from zero to one using the RAND() function of Excel. Every occurrence of a random number that is less than or equal to w/300 is marked as a win by white. Those greater than 1-w/300-b/300 are marked as a draw. The rest are marked as a win by black. This procedure is repeated one thousand times to generate one thousand simulated repetitions of the experiment. The variations in the P[$p_w$, $p_b$, $p_d$] vector among these simulated repetitions serves as our measure of uncertainty in our estimated value of π[$p_w$, $p_b$, $p_d$]. The computational details and simulation results for both Group 1 openings (Munshi, Group 1 Simulations, 2014) and Group 2 openings (Munshi, Group 2 Simulations, 2014) are available in the data archive for this paper.

To compare two openings, we plot the simulation results for both openings in the same Cartesian xy space with x= number of white wins and y=number of black wins as shown in Figure 1[8]. Each marker in the plot represents a unique simulated match of 300 games. In this particular example we note that the two openings appear to form distinct clusters of games. The visual comparison appears to indicate that white wins more games and black wins fewer games in the baseline opening than in the test opening. The essential research question is whether the unobservable probability vectors that generated the two sets of games are different; or whether both sets of simulated results could have been generated by the same underlying probability vector π[$p_w$, $p_b$, $p_d$]. If they are different, the difference is described as an effect of the opening since all other variables are controlled and fixed for both sets of games.

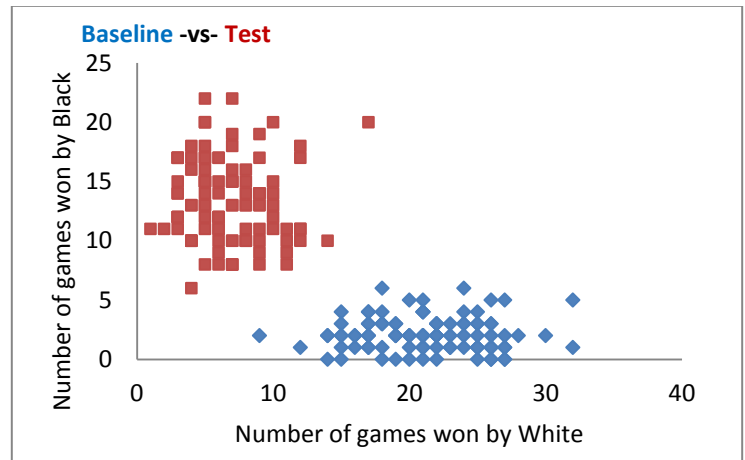

Figure 1 Graphical display of simulated match outcomes

### 3.3  Hypotheses:          The research question may be stated in hypothesis format as follows:

Ho: π[$p_w$, $p_b$, $p_d$](test opening) = π[$p_w$, $p_b$, $p_d$] (baseline opening)
Ha: π[$p_w$, $p_b$, $p_d$] (test opening) ≠ π[$p_w$, $p_b$, $p_d$] (baseline opening).

---

[8] In all such graphs presented in this paper the baseline control opening is represented by diamond markers and the test opening is represented by square markers.

In essence, if the same underlying probability vector could have generated both of the observed experimental results then we fail to reject Ho. Otherwise we reject Ho and conclude that there is an opening effect and use the graph to determine whether the innovation is a success or a failure and if it is a failure whether the failure is benign on injurious to the innovator. For example, if Ho is rejected in the case shown in Figure 1 and if the innovator is black, then the innovation is successful; but if the innovator is white then the innovation is a failure and injurious to the innovator. If we fail to reject Ho, we conclude that test opening is a benign innovation.

The hypothesis is tested by computing the observed Euclidean distance[9] between the centroids of the two treatment clusters and the standard deviation of the Euclidean distances of the simulated games from their respective centroids. We assume that if the two $\pi[p_w, p_b, p_d]$ vectors are the same then the distance between them is zero; and conversely, that if the distance between them is greater than zero, they must be different. We can then state our testable hypotheses as follows:

<div align="center">

Ho: distance between centroids = 0

Ha: distance between centroids $\neq$ 0

</div>

We test the hypothesis using the t-distribution holding our comparison error rate to $\alpha$=0.001 and our experiment-wide error rate to $\alpha$=0.005. These values of $\alpha$ have been proposed by Valen Johnson who makes a strong case that they enhance the reproducibility of results (Johnson, 2013). The Bonferroni adjustment for multiple comparisons (Abdi, 2007) implies that up to five comparisons may be made to each baseline opening experiment within an overall experiment-wide error rate of to $\alpha$=0.005.

## 4. DATA ANALYSIS AND RESULTS

Table 2 shows the data from twelve engine experiments. The essential data in the table are White = number of games won by white, Black = number of games won by black, Draw = number of games that ended in draw, and Total = the number of games played. The moves played in all 3,600 games have been made available in PGN format in the data archive for this paper (Munshi, OpeningPaperData, 2014).

| Expt# | Group | Description | White | Black | Draw | Total |
|---|---|---|---|---|---|---|
| 1 | 1 | **C68 Ruy Lopez** | 16 | 3 | 281 | 300 |
| 2 | 1 | C61 Birds Defense | 56 | 2 | 242 | 300 |
| 3 | 1 | C50 Giuoco Piano | 3 | 7 | 290 | 300 |
| 4 | 1 | C44 Scotch Game | 12 | 7 | 281 | 300 |
| 5 | 1 | C41 Philidor | 32 | 0 | 268 | 300 |
| 6 | 1 | C37 Kings Gambit | 4 | 29 | 267 | 300 |
| 7 | 2 | **B53 Sicilian Defense** | 21 | 2 | 277 | 300 |
| 8 | 2 | B21 Smith Morra | 7 | 13 | 280 | 300 |
| 9 | 2 | B22 Sicilian Alapin | 17 | 3 | 280 | 300 |
| 10 | 2 | B12 Caro Kann | 37 | 4 | 259 | 300 |
| 11 | 2 | B01 Scandinavian | 59 | 0 | 241 | 300 |
| 12 | 2 | B07 Pirc | 60 | 2 | 238 | 300 |

**Table 2 Observed sample statistics: raw data: UPDATED IN MARCH 2014**

[9] In all references to "distance" we mean the absolute value of the Euclidean distance.

**4.1    The baseline openings and the level of play.** Consider the for data for the two baseline openings shown in Table 2 as Experiment #1 and Experiment #7.  Note that more than 90% of the games in these experiments ends in draw, that white wins more games than black, and that black does not win more than 1% of the games. If we combine the two experiments into a 600-game sample we can estimate that the move imperfection rate is 1.67% and that white's first move advantage is 4.5%. The probability that the data are tainted by bad engine moves is therefore assumed to be low and inconsequential to the findings. These relationships are indicative of a high level of play and serve to validate the use of these openings as neutral and "perfect" baselines against which the test openings may be compared (Wikipedia, 2013).

The Monte Carlo simulations of the baseline experiments are depicted graphically in Figure 2. Each simulated experiment was generated with the estimated probability vectors $P[p_w, p_b, p_d]$ (Ruy Lopez) = $[0.0533, 0.01, 0.9367]$ and $P[p_w, p_b, p_d]$ (Sicilian) = $[0.07, 0.0067, 0.9233]$ and a sample size of 300 games. The graph in Figure 2 shows that the there are many overlapping results that could have been generated by either estimated probability vector and that the two openings do not form two distinct clusters of games. So we suspect that the two observed experiment results could have been generated by the same underlying and unobservable probability vector $\pi[p_w, p_b, p_d]$. In fact, the distance between the openings is quite small and the t-test shows a p-value of 0.254 much larger than our $\alpha = 0.001$. So we fail to reject Ho in this case and conclude that $\pi[p_w, p_b, p_d]$ (B53) could be equal to $\pi[p_w, p_b, p_d]$ (C68) because the evidence does not show that the two probability vectors are different. The two baseline openings are thus not only efficient but equivalent. The simulation data and computational details of the comparison are available in the online data archive for this paper (Munshi, C68 C53 Comparison, 2014).

The game data show that the engines have played generally the mainline moves one finds in grandmaster games (Jones & Powell, 2014). In the C68 Ruy Lopez experiment we find that all 300 games have maintained the C68 designation through the entire game. In the B53 Sicilian Defense experiment most of the games pertain to the B53 designation but B90 Sicilian Najdorf, B92 Sicilian Najdorf, B73 Sicilian Dragon, and B76 Sicilian Dragon lines also occur. In the rest of this study all references to the ECO designation B53 Sicilian Defense should be interpreted in this context. The actual moves played are shown in PGN format for both baseline openings, C68 Ruy Lopez (Munshi, Experiment 01, 2014) and C53 Sicilian Defense (Munshi, Experiment 07, 2014), in the data archive for this paper.

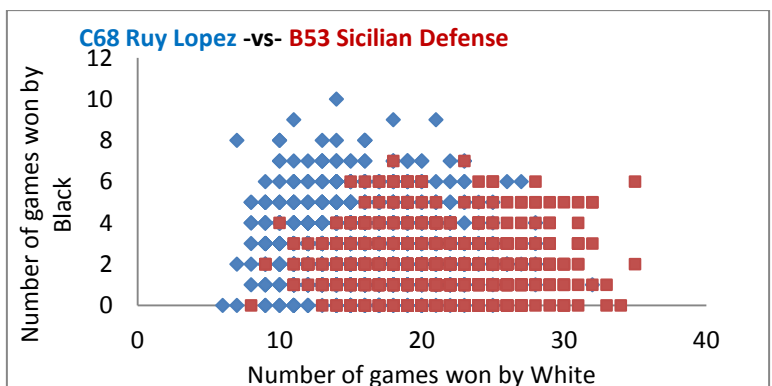

Figure 2 Monte Carlo Simulation of the baseline openings

**4.2    Hypothesis tests**.  Table 3 is a summary of the hypothesis tests. Ten hypothesis tests are made. The null hypothesis is rejected in six of them. We are now in a position to interpret the results for each test opening shown in Table 3 by examining the plot of the simulation data in light of the hypothesis test results. Our objective is to use this information to classify each test opening innovation into one of three categories listed below.

Category A:       The innovation has succeeded by changing the probability vector in favor of the innovator.
Category C:       The innovation is benign. The data do not indicate that it has changed the probability vector.
Category F:       The innovation has failed. It has changed the probability vector to the detriment of the innovator.

| Expt# | Group | Test Opening | Compare with | Distance | Stdev | t-value | p-value | α | Decision |
|---|---|---|---|---|---|---|---|---|---|
| 2 | 1 | C61 Birds Defence | C68 Ruy Lopez | 40.012 | 5.861 | 6.827 | 1.1E-11 | 0.001 | Reject Ho |
| 3 | 1 | C50 Giuoco Piano | C68 Ruy Lopez | 11.705 | 3.904 | 2.998 | 0.00275 | 0.001 | Fail to reject |
| 4 | 1 | C44 Scotch Game | C68 Ruy Lopez | 5.657 | 4.266 | 1.326 | 0.18497 | 0.001 | Fail to reject |
| 5 | 1 | C41 Philidor | C68 Ruy Lopez | 16.279 | 4.949 | 3.289 | 0.00102 | 0.001 | Fail to reject |
| 6 | 1 | C37 Kings Gambit | C68 Ruy Lopez | 28.636 | 5.016 | 5.709 | 1.3E-08 | 0.001 | Reject Ho |
| 8 | 2 | B21 Smith Morra | B53 Sicilian Defense | 17.804 | 4.508 | 3.949 | 8.1E-05 | 0.001 | Reject Ho |
| 9 | 2 | B22 Sicilian Alapin | B53 Sicilian Defense | 4.123 | 4.522 | 0.912 | 0.362 | 0.001 | Fail to reject |
| 10 | 2 | B12 Caro Kann | B53 Sicilian Defense | 16.125 | 5.367 | 3.005 | 0.0027 | 0.001 | Fail to reject |
| 11 | 2 | B01 Scandinavian | B53 Sicilian Defense | 38.053 | 5.839 | 6.517 | 9.1E-11 | 0.001 | Reject Ho |
| 12 | 2 | B07 Pirc | B53 Sicilian Defense | 39.000 | 5.848 | 6.669 | 3.3E-11 | 0.001 | Reject Ho |

Table 3 Hypothesis tests: UPDATED IN MARCH 2014

## 5. DISCUSSION OF RESULTS

### 5.1    Experiment #2: C61 Bird Defense: Category F

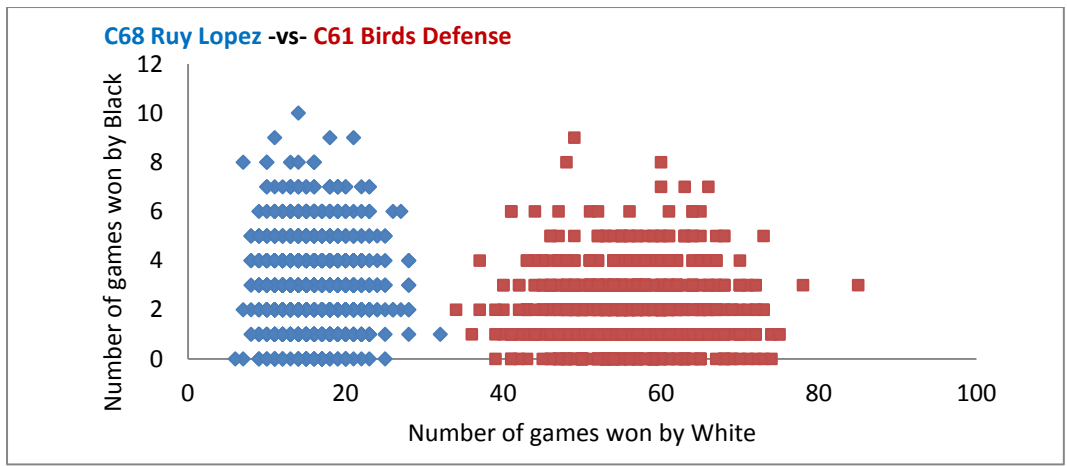

Figure 3 C61 Bird Defense

<u>Hypothesis test</u>:            In the hypothesis test in Table 3 we find that the observed distance between C61 and C68 in our sample is 40.012. Since Ho is rejected we can conclude that

$$\pi[p_w, p_b, p_d]C61) \neq \pi[p_w, p_b, p_d]\ (C68)$$

If the $\pi[p_w, p_b, p_d]$ vectors of the two openings were the same, the probability that we would observe a distance of 40.012 or more between them in our sample is very close to zero and much less than our threshold of disbelief.

<u>Graph interpretation</u>:    We now examine Figure 3 for clues to how these probability vectors might differ and we find that the probability vector that drives game outcomes in the Bird Defense must contain a higher probability of a win by white. Since the innovator is black, the innovation has failed. We therefore classify the opening as Category F. The computational details for the comparison of these two openings are available in the data archive for this paper (Munshi, C68 C61 Comparison, 2014).

<u>General opinion</u>:          The finding is consistent with the rarity of the Bird Defense at the highest levels of play (Jones & Powell, 2014). However, it is interesting to note that there has been no call for a refutation of the Bird Defense except for an obscure note by chess teacher Edward Scimia that after 3…Nd4, "white will emerge with a small advantage due to having a better pawn structure" (Scimia, 2013).

<u>Transpositions</u>:          There were no transpositions in this opening. The opening designation remained as ECO C61 Bird Defense in all 300 games. The PGN file is available in the data archive of this paper (Munshi, Experiment02, 2014).

## 5.2    Experiment #3: C50 Giuoco Piano: Category C

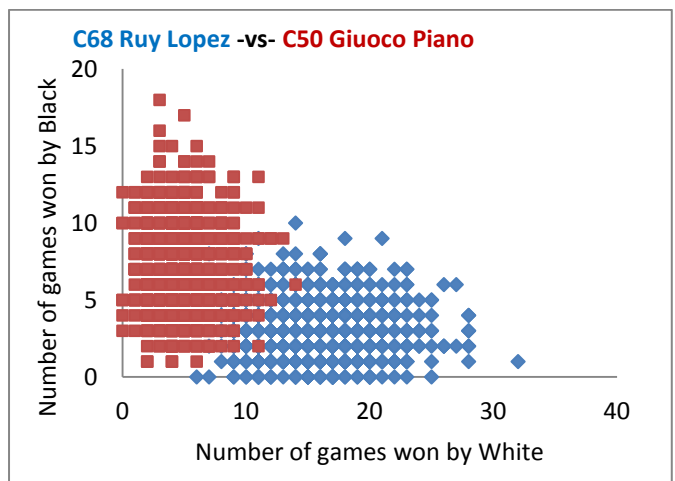

**Figure 4  C50 Giuoco Piano**

<u>Hypothesis test</u>:            In Table 3 we find no evidence that the Giuoco Piano opening is driven by a different probability vector than the one that generates games in the Ruy Lopez. If the $\pi[p_w, p_b, p_d]$vectors of the two openings were the same, the probability that we would observe a Euclidean distance of 11.705 or more between them in our sample is 0.00275, greater than our threshold of α=0.001. We therefore conclude that it is possible that chess games in the two openings may be driven by the same underlying probability vector.

Graph interpretation:    In Figure 4, we find that, even though the two simulation clusters appear to be different, there is an area of overlap between the two openings. This graph is consistent with the notion that the two sets of match outcomes could have been produced by the same underlying probability vector. Since we failed to reject Ho, we classify the innovation as Category C. It is a benign innovation. The computational details for this comparison is available in the data archive for this paper (Munshi, C68 C50 Comparison, 2014)

General opinion:       The finding is consistent with the general opinion of chess players. The Giuoco Piano is generally accepted as an equal to the Ruy Lopez (Kebu Chess, 2008). There has been no call for it to be refuted. Yet, the Giouoco Piano is rarely played by grandmasters (Jones & Powell, 2014). In my own collection of recent grandmaster games[10], the Giuoco Piano is not played at all (Munshi, GM 2012 2013 Book, 2014).

Transpositions:          Although the first three moves were set according to the ECO designation C50 Giuoco Piano, many of the games transposed into ECO code C54. The PGN file is available in the data archive of this paper (Munshi, Experiment03, 2014).

### 5.3    Experiment #4: C44 Scotch Game: Category C

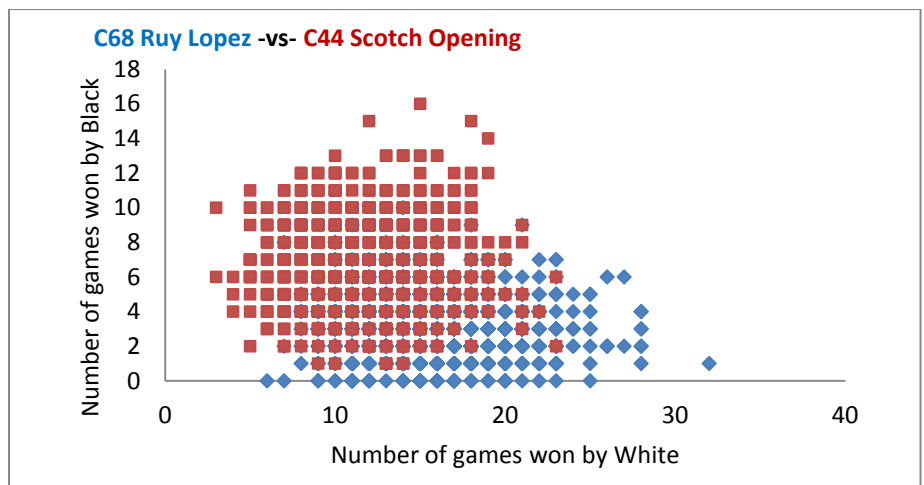

**Figure 5 C44 Scotch Game**

Hypothesis test:        In Table 3 we find that the distance between C44 and C68 is 5.657 and the probability that we would observe a distance this large or larger if if the π[$p_w$, $p_b$, $p_d$]vectors that generate game outcomes in these openings were the same is 0.18497, greater than our threshold of disbelief. We are therefore unable to reject the possibility that the game outcomes in the two openings are driven by the same underlying $\pi[p_w, p_b, p_d]$ vector.

---

[10] Unlike the online databases my game collection excludes blitz, rapid, and simul games.

Graph interpretation:     In Figure 5, we find that the simulated experiments do not form distinct clusters. There is a significant area of overlap. This graphic supports the notion that match outcomes in the two openings could have been generated by the same underlying probability vector. We therefore classify the innovation as Category C. It is a benign innovation. The computational details for the comparison of these two openings are available in the data archive for this paper (Munshi, C68 C44 Comparison, 2014).

General opinion:     The finding is consistent with the general opinion of chess players. The Scotch Game is generally accepted as a strong opening by grandmasters (Gserper, 2009) and there has been no call for it to be refuted. However, it is rarely seen in grandmaster tournaments. In opening books we find that at high levels of play 3. Bb5 is preferred 11:1 over 3. d4 in this line (Jones & Powell, 2014).

Transposition:     The first three moves in the experiment were set according to ECO code C44 but many of the games had transposed into C45. The games may be viewed in PGN format in the online data archive of this paper (Munshi, Experiment04, 2014).

## 5.4     Experiment #5: C41 Philidor:  Category C

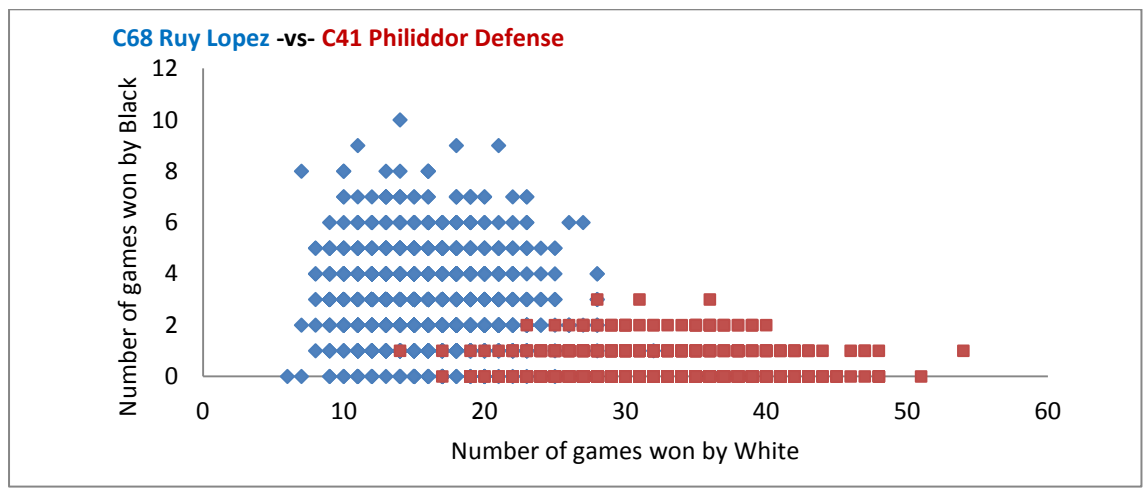

**Figure 6 C41 Philidor Defense**

Hypothesis test:     Although the distance between C41 Philidor and C68 Ruy Lopez is 16.279 it is still not large enough for it to be unusual even under the assumption that the same $\pi[p_w, p_b, p_d]$ vector is responsible for generating game outcomes in both openings because the p-value of 0.00104 is not less than our threshold of disbelief that is set to $\alpha=0.001$. Thus we must allow for the possibility that

$$\pi[p_w, p_b, p_d](C41) = \pi[p_w, p_b, p_d](C68).$$

Graph interpretation:     In Figure 6 it seems that the two game clusters are different but we find that there is an area of overlap that contain games from both openings. The graph is consistent with the inconclusive nature of our hypothesis test. Since the two results of these two experiments could have been generated by the same π[$p_w$, $p_b$, $p_d$] vector we classify the C44 Philidor as Category C. It is a benign innovation. The computational details of the comparison are included in the data archive for this this paper (Munshi, C68 C41 Comparison, 2014).

General opinion:        The finding is consistent with the general opinion of chess players. The Philidor Defense is generally accepted as a strong opening (Bauer, 2006) and there has been no call for it to be refuted. However, in opening books we find that 2…Nc6 is preferred to the 2…d6 innovation by a ratio of 100:1 at high levels of play (Jones & Powell, 2014). In my collection of recent high profile grandmaster tournaments the Philidor is notable in its absence.  (Munshi, GM 2012 2013 Book, 2014).

Transpositions:        The first three moves in the experiment were set according to ECO code C41and this ECO designation remained unchanged throughout all 300 games. The games may be viewed in PGN format in the online data archive of this paper (Munshi, Experiment 05, 2014).

## 5.5        Experiment 6: C37 King's Gambit: Category F

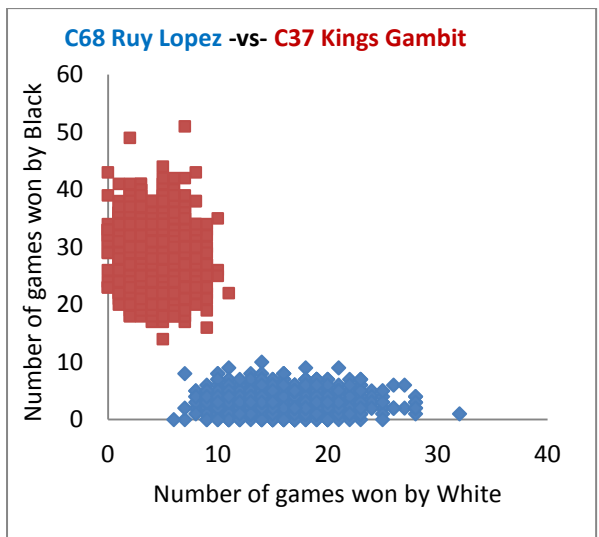

**Figure 7 C37 King's Gambit**

Hypothesis test:        In the hypothesis test in Table 3 we find that the observed distance between C37 and C68 in our sample is 28.636. Since Ho is rejected we can conclude that

$$\pi[p_w,\, p_b,\, p_d]\ (C37) \neq \pi[p_w,\, p_b,\, p_d]\ (C68)$$

If  the π[$p_w$, $p_b$, $p_d$] vectors of the two openings were the same, the probability that we would observe a distance of 28.636 or more between them in our sample is very close to zero and much less than our threshold of disbelief set to α=0.001.

Graph interpretation:        Figure 7 shows two distinct clusters of simulated match results that do not overlap. The graph visually supports our conclusion that the game outcomes in the two openings are driven by different $\pi[p_w, p_b, p_d]$ vectors. In particular, the 2. f4 innovation by white appears to have increased $p_b$ and/or decreased $p_w$ both of which work to the detriment of the innovator. We therefore classify the opening in Category F. It is a failed innovation. . The computational details of the comparison between C37 King's Gambit and C68 Ruy Lopez are available in the data archive for this paper (Munshi, C68 C37 Comparison, 2014).

General opinion:          Former world champion Bobby Fischer had once published a paper calling for the refutation of the King's Gambit (Fischer, 1961) and more recently chess engine programmer Vasik Rajlich carried out an extensive study on chess engines to support Fischer's call (Rajlich: Busting the King's Gambit, 2012). In opening books we find that 2. Nf3 is preferred to 2. f4 in this line by a ratio of 43:1 at high levels of play (Jones & Powell, 2014). Yet, the King's Gambit enjoys a degree of popularity and has many ardent supporters (Kebu Chess, 2008).

Transpositions:          The first three moves of the games were set according to ECO code C37 but many of the games in the sample transposed to C39. A complete record of all 300 games is available in PGN format in the data archive for this paper (Munshi, Experiment 06, 2014).

### 5.6      Experiment #8: B21 Smith Morra: Category F

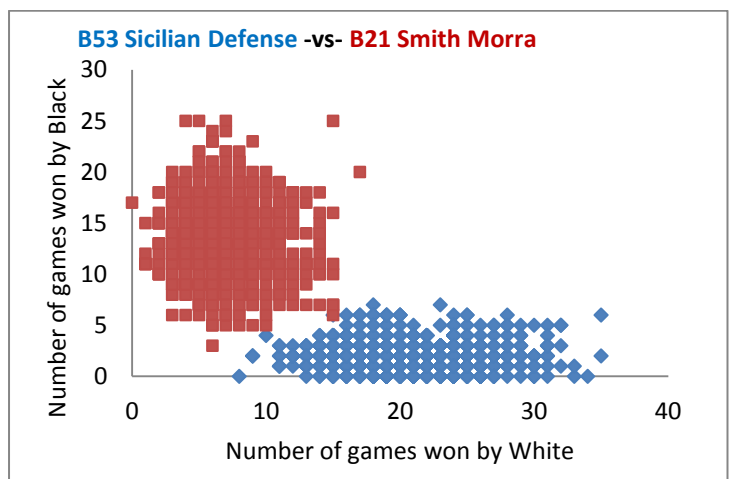

**Figure 8 B21 Smith Morra Gambit**

Hypothesis test:          In the hypothesis test in Table 3 we find that the observed distance between B21 and B53 in our sample is 17.804. Since Ho is rejected we can conclude that

$$\pi[p_w, p_b, p_d] \ (B21) \neq \pi[p_w, p_b, p_d] \ (B53)$$

If the $\pi[p_w, p_b, p_d]$ vectors of the two openings were the same, the probability that we would observe a distance of 17.804 or more between them in our sample is 0.000081, much less than our threshold of disbelief set to $\alpha=0.001$.

<u>Graph interpretation</u>:      Figure 8 shows two distinct clusters of simulation results that do not overlap. It visually supports our conclusion that the game outcomes in the two openings are driven by different $\pi[p_w, p_b, p_d]$ vectors. In particular, the 2. d4 innovation by white appears to have increased $p_b$ and/or decreased $p_w$ both of which work to the detriment of the innovator. We therefore classify the opening in Category F. It is a failed innovation. The computational details of the comparison between B21 and B53 are available in the data archive for this paper (Munshi, B53 B21 Comparison, 2014).

<u>General opinion</u>:        The opening continues to be controversial with no general consensus as to its merit. There are high profile commentators on both sides if the argument. For example Marc Esserman (Esserman, 2012) is a supporter while Timothy Taylor claims to have found its weakness (Taylor, 1993). In online opening databases we find that the Smith Morra is not played by grandmasters. At the highest levels of play, 2. Nf3 is preferred to 2. d4 in this line by a ratio of 1459:1 (Jones & Powell, 2014).

<u>Transpositions</u>:        There were no transpositions in this experiment. The ECO designation remained as B21 throughout every game. The PGN file is available in the data archive for this paper (Munshi, Experiment 08, 2014).

## 5.7    Experiment  #9: B22 Sicilian Alapin: Category C

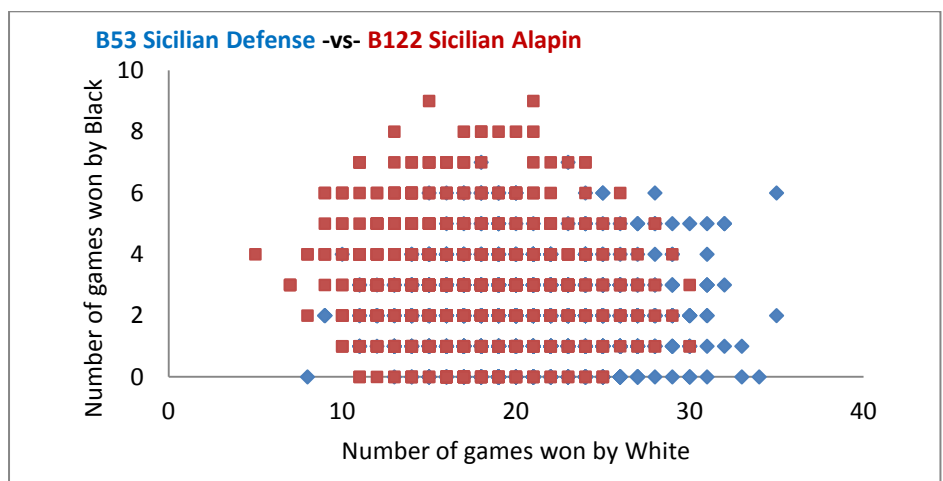

**Figure 9 B22 Sicilian Alapin**

<u>Hypothesis test</u>:        In Table 3 we find that the distance between B53 and B22 is 4.123 and the probability that we would observe a difference as large or larger is 0.362, much larger than our threshold of disbelief set at 0.001. Since we are unable to reject Ho, we must allow for the possibility that the same underlying $\pi[p_w, p_b, p_d]$ vector generates game outcomes for both of these openings.

<u>Graph interpretation</u>:    In Figure 9 we see the simulated matches overlie each other and form a single cluster. The graph supports the null hypothesis that a single underlying $\pi[p_w, p_b, p_d]$ vector could generate the observed results from both of these openings. We therefore classify the B22 Alapin in Category C. The computational details of the comparison is available in the data archive for this paper (Munshi, B53 B22 Comparison, 2014).

<u>General opinion</u>:        There does not exist any negative opinion on the B22 Alapin variation and there are many who feel that 2. c3 is a strong move by white in the very popular Sicilian opening (Eddleman, 2010). Yet it is a rare occurrence in grandmaster games. In online opening databases one finds that 2. Nf3 is preferred to 2. c3 by a ratio of 35:1 (Jones & Powell, 2014). Our finding seems to be in agreement with general opinion but at odds with the rarity of the 2. c3 move in this line.

<u>Transpositions</u>:        There were no transpositions. The ECO designation remained as B22 through all 300 games. The games may be viewed in PGN format online (Munshi, Experiment 09, 2014).

## 5.8        Experiment #10: B12 Caro-Kann: Category C[11]

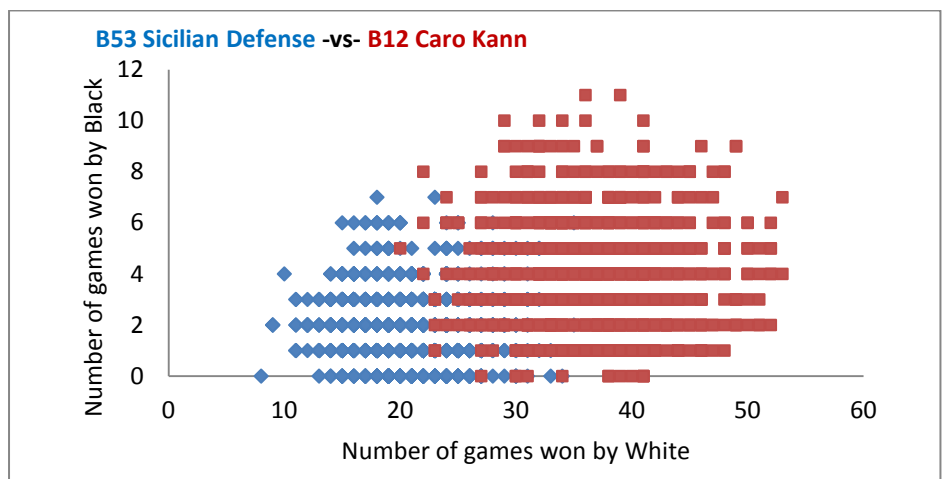

**Figure 10 B12 Caro-Kann: UPDATED IN MARCH 2014**

<u>Hypothesis test</u>:        In the hypothesis test in Table 3 we find that the observed distance between B12 and B53 in our sample is 16.125. Since we failed to reject Ho we must allow for the possibility that

$$\pi[p_w, p_b, p_d] \ (B12) = \pi[p_w, p_b, p_d] \ (B53)$$

If  the $\pi[p_w, p_b, p_d]$ vectors of the two openings were the same, the probability that we would observe a distance of 16.125 or more between these two openings in our sample is 0.0027. Since the probability is not less than our threshold of disbelief set to $\alpha$=0.001 we conclude that the same probability vector could have generated both B12 and B53 game outcomes.

<u>Graph interpretation</u>:        Figure 10 appears to show that the $\pi[p_w, p_b, p_d]$ (B12) vector contains higher win probabilities for both white and black but the degree of overlap between the two clusters supports our hypothesis test conclusion that there is no evidence here that game outcomes in the B12 is driven by a different probability vector for B53 . Computational details of the comparison between B12 Caro-Kann and B53 Sicilian Defense are included in the data archive for this paper (Munshi, B12 B53 Comparison, 2014)

---

[11] The Caro-Kann data were updated in March 2014 to correct a data entry error

<u>General opinion</u>:        There have been some calls for the refutation of this opening (KenilworthKibitzer, 2010) but most commentators and analysts consider the Caro-Kann to be a strong opening comparable to the B53 Sicilian (Kebu Chess, 2008). However, in online opening databases we find that at the grandmaster level,  1…c5 is preferred to 1…c6 by a ratio of only 4: 1 (Jones & Powell, 2014).

<u>Transpositions</u>:        There were no transpositions. The first three moves were specified according to the B12 Caro-Kann Advance variation and the opening designation stayed in that ECO code throughout the games. The line used represents the mainline in the opening book set to Elo ratings of 2600 or higher. The moves made in these games are included in the data archive for this paper (Munshi, Experiment 10, 2014).

### 5.9        Experiment #11: B01 Scandinavian: Category F

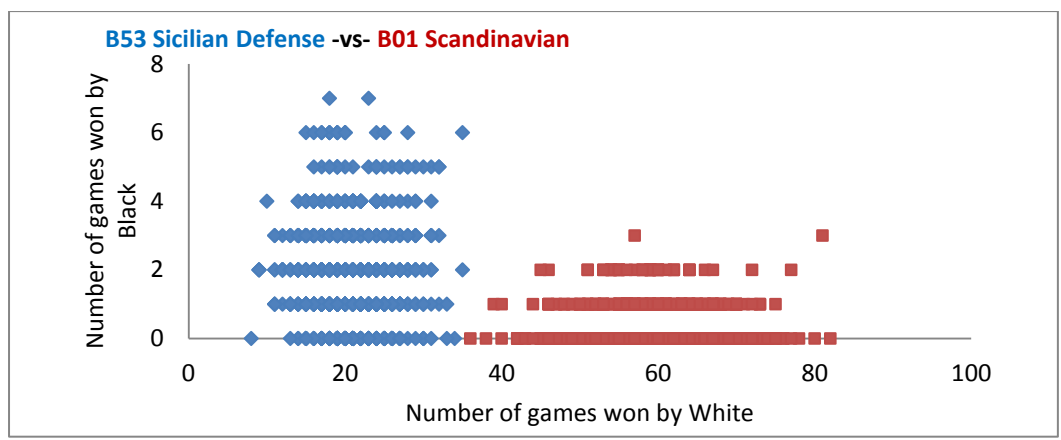

**Figure 11 B01 Scandinavian Defense**

<u>Hypothesis test</u>:        In Table 3 we find that the distance between B01 Scandinavian and B53 Sicilian is 38.053 and the probability that we would observe this difference or larger if game outcomes in the two openings were driven by the same probability vector is very close to zero and much less than our threshold value of 0.001. We therefore conclude that

$$\pi[p_w, p_b, p_d] \ (B01) \neq \pi[p_w, p_b, p_d] \ (B53)$$

<u>Graph interpretation</u>:        Figure 11 clearly shows that the two clusters of game outcomes are distinct and separated with no overlap,  in agreement with our hypothesis test. The graph appears to show that the B01 innovation of 1…d5 has increased $p_w$ or decreased $p_b$ or both. As these changes work against the innovator we classify this opening as Category F. Computational details of the comparison of B01 with B53 are included in the data archive for this paper (Munshi, B01 B53 Comparison, 2014).

<u>General opinion</u>:        There is plenty of support for this opening even at the grandmaster level of chess (Gserper, Scandinavian center counter defense, 2010). There has been no call for refutation. In opening books we find that 1…c5 is preferred to 1…d5 by a ratio of 39:1 (Jones & Powell, 2014). Our finding is at odds with subjective opinion but consistent with its rarity at a high level of play.

          There were no transpositions. The ECO code remained as B01 throughout every game played. The games may be viewed in the data archive for this paper (Munshi, Experiment 11, 2014).

### 5.10    B07 Pirc Defense: Category F

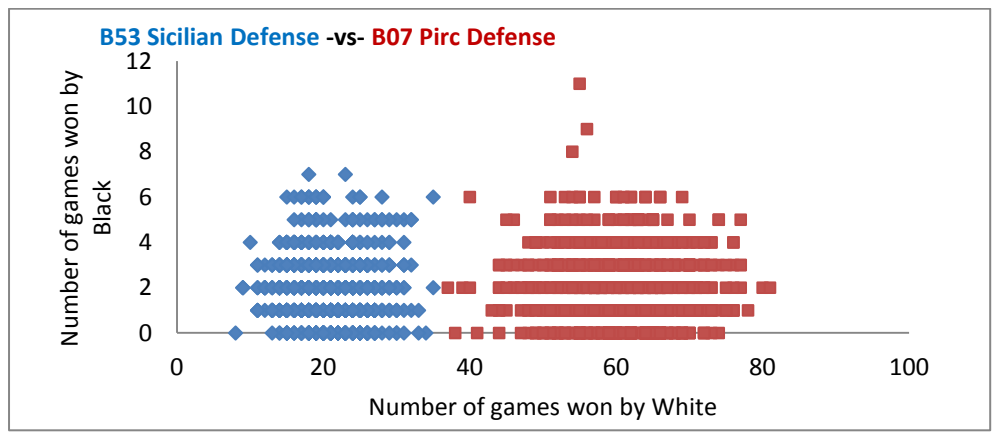

**Figure 12 B07 Pirc Defense**

Hypothesis test:          Table 3 shows that the distance between B07 and B53 is 39 and the probability of observing a distance this large or larger if $\pi[p_w, p_b, p_d]$ (B07) = $\pi[p_w, p_b, p_d]$ (B53) is close to zero and less than our threshold of 0.001. We therefore reject Ho and conclude that the data provide strong evidence that

$$\pi[p_w, p_b, p_d] \text{ (B07)} \neq \pi[p_w, p_b, p_d] \text{ (B53)}$$

and that the chess games using these two openings are therefore driven by different probability vectors.

Graph interpretation:    The simulated matches in Figure 12 form distinct clusters separated from each other without any overlapping match outcomes. This graphic serves to visually confirm out conclusion in the hypothesis test. It is also evident in the graph that the difference between the two probability vectors may include a higher value of $p_w$ in B07 than in B53. This difference works against the black who is the innovator in this case. We therefore classify the 1…d6 innovation as Category F. The computational details of the comparison are included in the data archive for this paper (Munshi, B53 B07 Comparison, 2014).

General Opinion:          The results conform with the relative rarity of this opening. At Elo ratings greater than 2600, the opening database shows that 1…c5 is preferred to 1…d6 by a ratio of over 18:1 (Jones & Powell, 2014). However there have not been a negative opinion expressed on this opening. Many experts and analysts have promoted the opening as one that opens up opportunities for black (Kebu Chess, 2012). (This paragraph was added in March 2014)

Transpositions:          The first three moves of the games were set according to B07 but many of the games transposed into B08 and A43. The complete game record move by move is included in the data archive for this paper (Munshi, Experiment 12, 2014).

## 6. CONCLUSIONS AND IMPLICATIONS[12]

Starting with a trinomial stochastic model for chess game outcomes we designed engine experiments and a Monte Carlo simulation procedure to detect the effect of the opening on the probability vector that generates game outcomes. The results show that the method of detecting opening effects is able to discriminate between known strong openings and known weak openings. Of the ten opening innovations tested, five were found to be failures because the innovation did not strengthen but in fact weakened the position of the innovator. The other five were found to be benign innovations as they had no measurable effect on the probability vector that generates game outcomes. None of the innovations tested succeeded in improving the innovator's chances of winning, leaving the baseline control openings as the strongest openings examined in this study.

The value of this work lies not so much in these findings but in offering an objective methodology for comparing chess openings. The method may be used as an analysis tool to compare complex variations in opening lines. It may also be used to test controversial openings for possible refutation and in those cases the method serves as an objective refutation tool.

An important implication of this work is that, because chess is trinomial and not binomial, neither opening strategies nor chess players may be compared using only a single scalar measure. The search for a single scalar index such as match score difference and Elo differential to measure relative playing strength ignores the three-dimensional nature of the probability vector that generates chess game outcomes. For example, in this paper we looked at both distance and direction. Although no opening in this study exhibited this behavior it is possible that an innovation that is rejected by the hypothesis test because of a large distance could still receive a grade of C if $p_w$ and $p_b$ are changed in the same proportion. In such a case, the innovation would serve to change only the probability of decisive games but without a relative advantage to either side. A single scalar measure of strength differential will never be found because the idea is mathematically flawed.

A weakness in the methodology is the possibility that the results may to some extent be an artifact of the engine since the same engine with the same parameters was used to play both sides of the board for every game in every experiment. Further research is under way to test this hypothesis. A weakness in the findings of this work that requires further investigation is that there are some anomalies between the rarity data shown in Table 1 and the p-values in Table 3. Although there are some spectacular agreements, the overall lack of correlation between these values requires an explanation. For example, the Sicilian Alapin, an opening that appears to be as strong as the Sicilian mainline is relatively unpopular; while openings graded as "F" in this study such as the Scandinavian and the King's Gambit are played by grandmasters and even promoted by analysts (Kebu Chess, 2012). Also, as in all experimental studies, we may have gained precision possibly at the expense of realism. The objective is the development of an  objective measure for comparing openings that may serve to bring many endless and subjective debates to a satisfactory conclusion and help to refine the opening book.

---

[12] Edited in March 2014